# Mapping the self-generated magnetic fields due to thermal Weibel instability

Chaojie Zhang[1, *], Yipeng Wu[1, †], Mitchell Sinclair[1], Audrey Farrell[1], Kenneth A. Marsh[1], Irina Petrushina[2], Navid Vafaei-Najafabadi[2,3], Apurva Gaikwad[2], Rotem Kupfer[3], Karl Kusche[3], Mikhail Fedurin[3], Igor Pogorelsky[3], Mikhail Polyanskiy[3], Chen-Kang Huang[4], Jianfei Hua[5], Wei Lu[5], Warren B. Mori[1,6], Chan Joshi[1, ‡]

[1] *Department of Electrical and Computer Engineering, University of California Los Angeles, Los Angeles, CA 90095, USA*

[2] *Department of Physics and Astronomy, Stony Brook University, New York, NY 11794, USA*

[3] *Accelerator Test Facility, Brookhaven National Laboratory, Upton, NY 11973, USA*

[4] *Institute of Atomic and Molecular Sciences, Academia Sinica, Taipei, 10617, Taiwan*

[5] *Department of Engineering Physics, Tsinghua University, Beijing 100084, China*

[6] *Department of Physics and Astronomy, University of California Los Angeles, Los Angeles, CA 90095, USA*

## Abstract

Weibel-type instability can self-generate and amplify magnetic fields in both space and laboratory plasmas with temperature anisotropy. The electron Weibel instability has generally proven more challenging to measure than its ion counterpart owing to the much smaller inertia of electrons, resulting in a faster growth rate and smaller characteristic wavelength. Here, we have probed the evolution of the two-dimensional distribution of the magnetic field components and the current density due to electron Weibel instability, in $CO_2$-ionized hydrogen gas (plasma) with picosecond resolution using a relativistic electron beam. We find that the wavenumber spectra of the magnetic fields are initially broad but eventually shrink to a narrow spectrum representing the dominant quasi-single mode. The measured $k$-resolved growth rates of the instability validate kinetic theory. Concurrently, self-organization of microscopic plasma currents is observed to amplify the current modulation magnitude that converts up to ~1% of the plasma thermal energy into magnetic energy.

## Introduction

A well-accepted hypothesis explaining micro-gauss magnetic fields observed throughout the cosmos is that the initially weak seed fields are amplified by the galactic dynamo[1,2]. The origin of such seeds, however, is still an open question in plasma astrophysics[3–5]. One candidate is the Biermann battery mechanism that generates magnetic fields in plasmas with non-parallel density and temperature gradients[6–8]. Another is the Weibel instability that is ubiquitous in plasmas with temperature anisotropy[9–15]. First theorized six decades ago, Weibel instability is arguably the earliest discovered plasma kinetic instability and yet its conclusive experimental verification has proven to be very challenging for two primary reasons. First, until recently it had not been possible to generate a plasma with a known temperature anisotropy as initially envisioned by Weibel[9]. Second, there was no suitable technique to measure the complex and evolving topology of the ultrafast (e.g., on the picosecond time scale in laser-produced plasmas) magnetic fields generated in the plasma. This is particularly the case for the electron Weibel instability because of its generally faster growth rate and shorter wavelength originating from the much smaller inertia of electrons compared to ions.

Since its discovery, extensive theoretical studies and particle-in-cell (PIC) simulations have revealed important characteristics of Weibel magnetic fields[16–18]. For instance, it is predicted that the Weibel instability in an infinite plasma initially grows with a broad $k$-spectrum, $0 < k < \sqrt{A}\omega_p/c$, where $k$ is the wavenumber of the magnetic field, $A \equiv \frac{T_{\text{hot}}}{T_{\text{cold}}} - 1$ is the temperature anisotropy of the plasma, $\omega_p$ the plasma frequency and $c$ the speed of light in vacuum. $T_{\text{hot}}$ and $T_{\text{cold}}$ are temperatures in the two orthogonal directions. The broad spectrum implies that many modes are excited simultaneously- each mode having an effective growth rate. As the temperature anisotropy decreases, the $k$ spectrum is expected to shrink to a narrow peak[18]. Physically, the narrowing of the $k$ spectrum is caused by the coalescence of plasma currents accompanied by the amplification of the magnetic fields. Once the quasi-single mode is formed, the magnetic field can maintain its topological structure for many plasma periods ($\gg \omega_p^{-1}$). Although most theoretical treatments assume a temperature anisotropy in two dimensions, plasma produced by the electric field of an ultrashort laser pulse may have different temperatures along all three Cartesian coordinates, which leads to the growth of multi-dimensional Weibel instability. It is crucial to experimentally demonstrate this evolution of the $k$-spectrum of the magnetic field to conclusively attribute it to the Weibel instability. To our knowledge this has not been done before.

It is well known that numerous kinetic instabilities- so called because they arise due to non-thermal (non-Maxwellian) and/or anisotropic velocity distribution functions of electrons and/or ions- can be self-excited in plasmas. These instabilities can be predominantly electrostatic/ longitudinal, electromagnetic/transverse, or a mixture of the two. We have previously shown that optical-field ionized (OFI) plasmas produced by intense femtosecond laser pulses are an excellent platform for studying this class of instabilities because uniform underdense plasmas with extremely high ($A \gg 10$) and predictable temperature anisotropies can be formed during the ionization process itself[19–21]. In previous work, we have documented the onset, saturation, and evolution of the streaming (electrostatic) and oblique current filamentation (electromagnetic with electrostatic components) instabilities by measuring the density fluctuations associated with them by Thomson scattering using a femtosecond probe pulse[20]. Furthermore, in a recent experiment, we have confirmed the existence of the quasi-static magnetic field directly by measuring the deflection of a relativistic electron probe beam[22]. The final topology of the magnetic field was consistent with that due to the Weibel instability. However, in that experiment, measurement of the onset and the subsequent evolution of the $k$-spectrum of the magnetic field, required to definitively attribute the phenomenon as a manifestation of the Weibel instability, was not possible.

Other previous work on the Weibel instability has involved the streaming of two counterpropagating laser-produced solid target plasmas[12–14]. In such cases the Weibel instability arises from the filamentation of the ions on a much longer timescale (typically on nanosecond timescale). One could call this a Weibel-type instability[10] because such a configuration too gives rise to a filamentary magnetic field in the overlap region of the two plasmas but it is not the scenario envisioned by Weibel, where the instability arises from the temperature anisotropy of a stationary plasma[9]. Attempts to control beam-plasma parameters more precisely were made in experiments where well-characterized electron beams from linear accelerators were sent through plasmas to study the relativistic current filamentation instability (CFI)[23]. The electron bunch was observed to form filamentary structures after its passage through the plasma. Similar behaviors of the electron bunch were also observed in experiments using ultrashort electron bunches from laser wakefield accelerators[24]. The study of CFI has also been extended to the overdense plasma regime in relativistic laser- or beam-solid interactions[25–30]. Recent studies show that in the case of dilute beam, the dominant mode is oblique[17,31].

In this article, we show the measurements of Weibel-generated magnetic fields in plasmas produced by ultrashort but intense, linearly polarized $CO_2$ laser pulses via optical-field ionization. We begin by producing anisotropic OFI hydrogen plasmas by using picosecond $CO_2$ laser pulses and then using ultrashort relativistic electron bunches from a linear accelerator to probe the magnetic fields. A movie of the magnetic fields with an exposure of ~ps and frame separation from a few to a few tens of ps was made using the electron beam probe. Analysis of the individual frames reveals how the $k$-spectrum and magnitude of the magnetic fields evolve as a function of time, with micron-level spatial resolution and thus allow us to validate the predictions of the Weibel theory.

## Results

### *OFI plasma with tri-Maxwellian electron velocity distribution (EVD)*

The key to our experiment is the initialization of known and highly anisotropic EVDs when the plasma is formed[19]. An example of the simulated EVDs in all three directions produced by ionization of hydrogen gas with an atomic density of $10^{18}$ cm$^{-3}$ by a 2-ps linearly polarized $CO_2$ laser is shown in Fig. 1. This example is from a three-dimensional (3D) particle-in-cell (PIC) simulation that mimics the experiment. The peak normalized vector potential in the simulation is $a_0 \equiv \frac{eA}{m_e c^2} \approx 0.12$ which is larger than $a_0 \approx 0.08$ that corresponds to the ionization threshold intensity $I_{th} \sim 10^{14}$ W/cm$^2$. Here the use of PIC simulation is essential to accurately predict the EVDs because the laser pulse duration is much longer than the plasma response time ($\omega_p^{-1} \sim 18$ fs for the plasma density $n_e = 10^{18}$ cm$^{-3}$) and therefore both the ionization process and the subsequent motion of electrons inside the laser and plasma fields need to be self-consistently modeled. In Fig. 1a we show the distribution of electrons in the 3D momentum space right after the laser has passed. The EVDs in each direction are plotted in Fig. 1b. In all three directions, the EVDs can be well fitted by a Maxwellian distribution. As expected, the plasma temperature in the laser polarization ($z$) direction is the highest ($T_z \approx 150$ eV). The temperature in the $y$ direction is the lowest ($T_y \approx 30$ eV), whereas the temperature in the laser propagation direction is higher ($T_x \approx 80$ eV). We note that if only the ionization process is considered, $T_x$ and $T_y$ should be similar. The increased $T_x$ indicates that the plasma has been preferentially heated in the longitudinal direction. Plasmas with such temperature anisotropy are unstable to Weibel instability as we will see below.

### *Deflection of the probe beam by self-generated magnetic fields in the plasma*

We performed an experiment at the Accelerator Test Facility of Brookhaven National Laboratory (ATF-BNL), and the layout of which is sketched in Fig. 1c. Anisotropic underdense plasmas [$n_e \approx (1.8 \pm 0.2) \times 10^{18}$ cm$^{-3}$] were created by ionizing a supersonic hydrogen gas jet using 2-ps, high-power (sub-terawatt) $CO_2$ laser pulses[32] (see Methods). The magnetic field in the plasma and its spatiotemporal evolution were probed by ultrashort relativistic electron bunches delivered by the ATF linear accelerator[33]. A set of permanent magnet quadrupoles (PMQs[34,35]) were used to relay and magnify the electron probe to a scintillator screen which converted the modulated electron flux to an optical image. With the movable PMQs inserted the image was magnified by 3.4x (further magnified by the optical system to an overall magnification of ~7.8x) with a spatial resolution of 2.9 $\mu m$.

By changing the delay of the electron probe with respect to the $CO_2$ laser, a movie of the density bunching of the electron beam due to its deflections by the magnetic field in the plasma was made (see Supplemental Materials). In Fig. 1d we show the raw data of the density modulations on the electron beam at representative times with respect to the plasma formation. The total time interval covered by these frames is ~150 ps. The images were obtained with PMQs inserted, with their object plane $10 \pm 0.5$ mm downstream of the plasma. The time zero was defined when the observable structures within the electron beam reached roughly the center of the field of view. In the next frame (3.3 ps later), the front of the structure has moved towards the right by ~1 mm as expected. The jitter between the laser and the electron probe was determined to be ~0.4 ps (see Supplemental Materials).

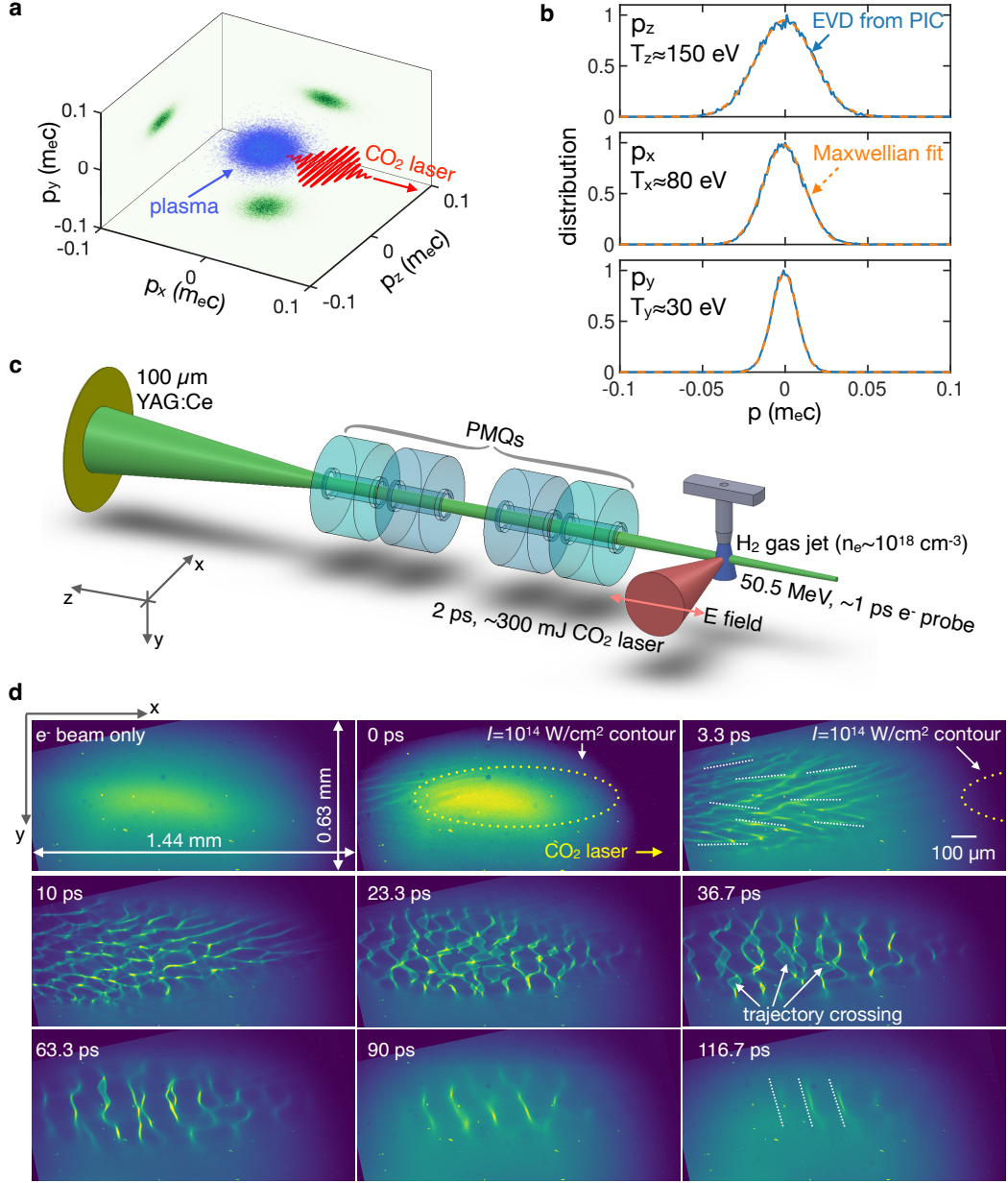

Fig. 1. EVD of the OFI plasma, experimental setup, and representative frames. (a) Simulated distribution of OFI plasma electrons in the 3D momentum space. (b) Projected EVDs (blue) and Maxwellian fits (red). In all three cases, the R-squared values are larger than 0.997. See Supplemental Material for the fitting residuals. These fits give the temperature anisotropies $A_{zy} = \frac{T_z}{T_y} - 1 \approx 4$ and $A_{zx} = \frac{T_z}{T_x} - 1 \approx 0.9$. (c) Sketch of the experimental layout. (d) Representative frames from the movie of the electron beam deflection by fields in the plasma. The first frame shows the e⁻ beam profile with no laser. The following frames show the evolution of the self-generated fields in the plasma. The yellow dotted ellipse on the 0 ps frame outlines the estimated $10^{14}$ W/cm² (ionization threshold) intensity contour of the $CO_2$ laser. The dotted white lines on the 3.3 ps and 116.7 ps frames are added to highlight the orientation of selected density strips. On the 36.7 ps frame, the white arrows mark structures caused by the trajectory crossing of the probe electrons which shifts the effective object plane closer to the plasma. All images were rotated counter-clockwise by 12 degrees to correct the PMQ-induced slant and put the longer dimension of the elliptical plasma parallel to the laser propagation direction.

We argue that the observed structures in the probe electron beam are caused by Weibel-generated magnetic fields in the plasma. In the experiment, the electron probe was orthogonal to the $CO_2$ laser pulse, i.e., in the positive $z$ direction in Fig.1c. Because of this probing geometry, the probe was deflected by the $B_x$ and $B_y$ components by the $\boldsymbol{v} \times \mathbf{B}$ force, not $B_z$. The contributions of small-amplitude wakes, stimulated Raman back scattering generated plasma waves[36], and other stochastic electric fields in the plasma are negligible because these fields oscillate at plasma frequency $\omega_p$ and therefore will be averaged over many periods ($\omega_p^{-1} \approx 13$ fs for $n_p = 1.8 \times 10^{18}$ cm$^{-3}$) as the picosecond long probe traverses the plasma[37,38]. In the 3.3 ps frame, the most prominent features seen are density strips within the electron beam that are approximately parallel to the laser propagation direction (see the dotted white lines in Fig. 1d). We will see in the next section that these electron density modulations can be transformed into magnetic fields and plasma current density maps. These horizontal density strips must arise because the probe electrons are deflected in the vertical ($y$) direction, which implies that within this period the dominant component of the magnetic field is $B_x$ with its wavevector along the $y$ direction. At this time, the $CO_2$ laser pulse has fully traversed the region probed by the electron beam, and the density structures have appeared in at least 2/3 of the frame. The evolution of the density structures in the $xy$-plane is seen in the back of the frame. The 10-ps frame is particularly interesting as it clearly shows that these initially predominantly horizontal strips (the rightmost side of the frame) start to break up in the $xy$-plane into smaller-scale "fish-net" structures on the left-hand side within ~1 ps. These "net" structures are in sharp focus in the 10-ps frame, which is indicative of deflected electrons coming to a focus at the object plane of the PMQs. A "net" structure means that the electrons are bent in both directions ($x$ and $y$) and that the two components $B_x$ and $B_y$ have approximately equal magnitude. The magnetic fields continue to grow, moving the object plane closer to the plasma, causing trajectory crossing of the probe electrons before they reach the object plane of the PMQs and blurring the structures as seen in the 23.3 ps frame. These "net" structures last for approximately 20 ps. In the next frame (36.7 ps), the density strips begin to line up in the vertical direction which indicates that the electrons are now predominantly deflected along the horizontal ($x$) direction by the $B_y$ component. As the instability continues to evolve, the density strips remain along the vertical direction but the spacing between strips keeps increasing and the structure appears to become a quasi-single mode. In the last frame (116.7 ps), the field has evolved to a quasi-single mode with a wavelength of ~145 μm (see dotted white lines). In addition to this morphological change, the magnitude of the electron probe density modulation also evolves as a function of time, which correlates with the evolution of the magnetic field amplitude.

### *Retrieved magnetic fields and current density distribution*

The path integrals of the magnetic fields along the probe propagation direction were retrieved by solving an equivalent optimal transport problem[39] (see Methods). The retrieved $\int B_x dz$ and $\int B_y dz$ fields for a representative frame (the 10-ps frame in Fig. 1d) are shown in Fig. 2a and b, respectively. The calculated path-integrated magnetic fields have a peak magnitude of ~100 Tesla×μm. For simplicity, the plasma was approximated as a slab (rather than the more realistic cylindrical plasma) with a thickness of 300 μm inferred from the transverse extent in the vertical direction. The retrieved path-integrated $B_x$ and $B_y$ fields have similar peak magnitudes of ~0.35 Tesla. Implications of violating this assumption are discussed later.

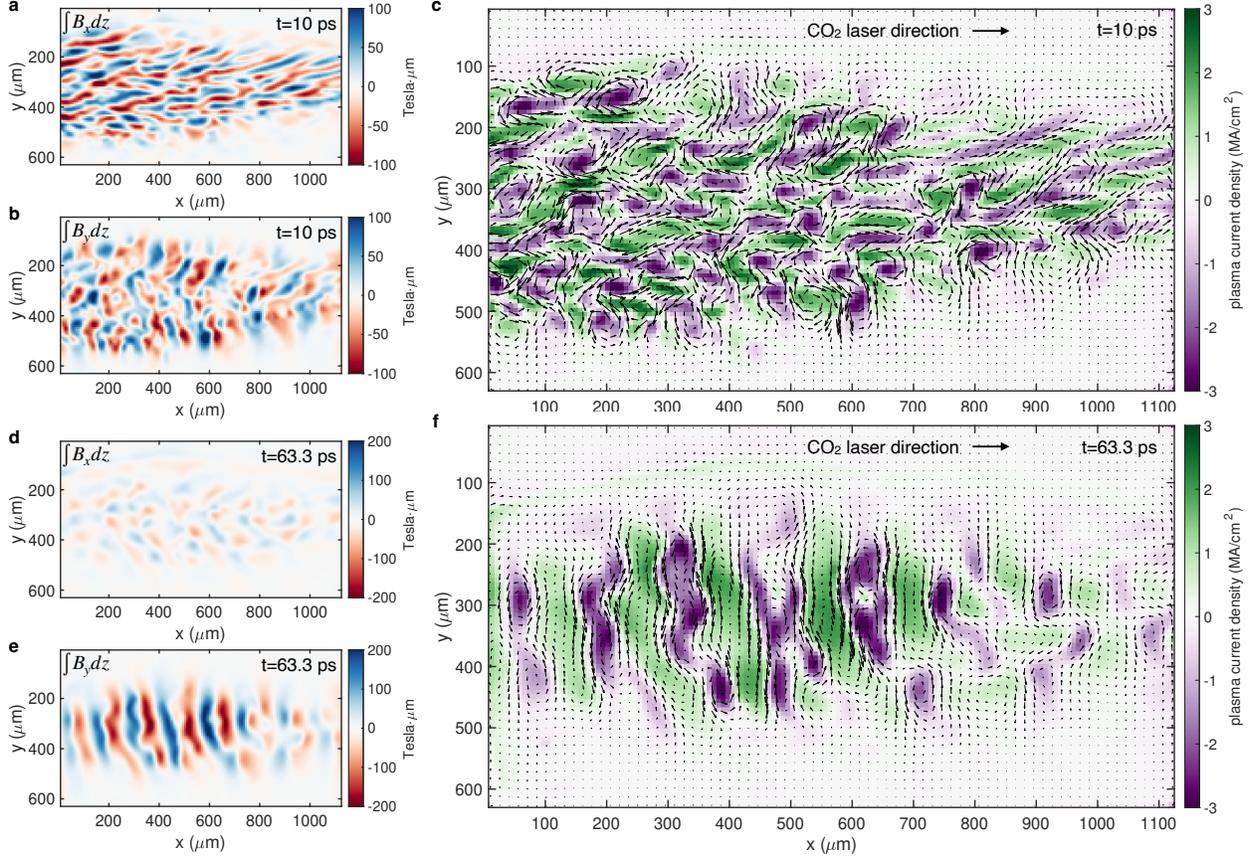

Fig. 2. Retrieved magnetic fields and current density maps. (a) and (b) show $\int B_x dz$ and $\int B_y dz$, for the 10 ps frame. The current density calculated using these magnetic fields is shown in (c) by the color scale whereas the arrows show the vector magnetic field $\mathbf{B_\perp} = \hat{x}B_x + \hat{y}B_y$. The corresponding results for the 63.3 ps frame are shown in (d)-(f).

The measured magnetic fields are predominantly generated by the quasi-static plasma currents, i.e., the contribution of the displacement current is small. Since the plasma temperature was the highest in the $z$ direction we anticipate that $J_z$ is the dominating source for the observed magnetic fields. Therefore, we calculate the current density $J_z$ using the retrieved two orthogonal magnetic field components by solving Maxwell's curl $\mathbf{H}$ equation (Ampere's law), $J_z = \mu_0^{-1}(\partial_x B_y - \partial_y B_x)$. The calculated 2D distribution of $J_z$ is shown in Fig. 2c by the color scale. The plasma current density is modulated along both $x$ and $y$ directions which supports the multi-dimensional nature of the instability.

The plasma currents keep merging as the instability grows. To highlight this, the magnetic fields and current density for the 63 ps frame are plotted in Fig. 2d-f. At this stage, the instability has saturated and the $B_y$ field dominates over $B_x$ (see Figs. 2d and e). Additionally, both the size of the current filaments and the spacing between them have increased. The morphology of the current distribution has changed from a 2D "fish-net" structure to a 1D periodic distribution along the $x$ direction, which is consistent with the $B_y$ field and the observed vertical density strips shown in Figs. 2e and 1d, respectively.

Using the retrieved current density, we can estimate the magnitude of the plasma current density modulation. Consider a plasma with Maxwellian EVD along the $z$ direction. It contains an equal number of electrons moving along opposing directions (e.g., $\pm z$). The initially unperturbed plasma current densities are $J_\pm(x) = \pm J_0$ where $J_0 = -en\langle v_{+z}\rangle = -env_{th}/\sqrt{2\pi}$ and is uniform on a spatial scale larger than the Debye length. Therefore, without modulation the net current density $J_z = J_+ + J_-$ is zero everywhere. The growth of Weibel instability causes coalescence of the microscopic plasma currents and leads to the spatial modulation of the current density, which can be assumed as $J_\pm = \pm J_0(1 \pm \frac{\epsilon}{2}\sin kx)$. Here $\epsilon/2$ and $k$ are the magnitude and wavevector of the modulation for $J_\pm$, respectively. This gives a net current density distribution of $J_z = J_0\epsilon\sin kx$. Using the experimental plasma density $n_e = (1.8 \pm 0.2) \times 10^{18}$ cm$^{-3}$ and $T_z \approx 150$ eV from the 3D PIC simulation, the magnitude of the net current density is calculated to be $59\epsilon$ MA/cm$^2$. The measured peak magnitude of $J_z$ is about 3 MA/cm$^2$ (see Figs. 2c and f), which corresponds to a modulation magnitude of $\epsilon \approx 5\%$. Note that in Weibel's theory, this current density modulation is caused by redistribution of the microscopic plasma currents and does not require plasma density modulation. Nevertheless, density modulation may emerge in the nonlinear stage of the instability[40]. We note that in a recent experiment, a current density modulation approaching unity in ion current filaments was measured using optical Thomson scattering[41].

The plasma beta is defined as $\beta \equiv \frac{p_{th}}{p_{mag}}$ where $p_{th} = nk_BT$ is the thermal pressure of the plasma and $p_{mag} = B^2/2\mu_0$ is the magnetic pressure. For $k_BT \approx 150$ eV, $n = 1.8 \times 10^{18}$ cm$^{-3}$ and the measured $\langle B^2\rangle^{1/2} \approx 0.35$ Tesla, these pressures are calculated as $p_{th} \approx 430$ bar and $p_{mag} \approx 0.5$ bar. Therefore, the plasma beta is $\beta \approx 860$. We note that this number represents the upper limit of the plasma beta in the experiment due to two reasons. First, there are structures indicating probe trajectory crossing after 23 ps (see Fig. 1d), which will reduce the retrieved magnitude of the magnetic field. In other words, the actual magnetic field strength is larger than the deduced one, by an estimated factor of ~2 which will reduce the $\beta$ by a factor of 4. Second, the $B$ field may not be uniform across the whole plasma along the probe propagation direction as assumed, namely, $B = \langle B \rangle + \delta B$, where $\langle B \rangle$ represents the average of $B$. The variation $\delta B$ leads to $\langle B^2 \rangle = \langle B \rangle^2 + \langle \delta B^2 \rangle = 2\langle B \rangle^2$ if we assume $\delta B \sim \langle B \rangle$. This means the actual plasma beta is two times smaller than that evaluated using $\langle B \rangle$. Taking both factors into account, the plasma beta may be reduced to ~100. This means that upon saturation, about 1% of the thermal energy in the plasma is converted to the magnetic field energy, which is on the same order (~4%) as observed in previous 3D PIC simulations with much hotter temperatures ($T_\perp = 16$ keV and $T_\parallel = 0.64$ keV)[18]. Similar level of energy conversion was also observed in PIC simulations of expanding plasmas[42] and in that work the authors suggest that conversion of a few percent of thermal energy into magnetic field energy via the Weibel instability may be sufficient for seeding the galactic dynamo[43,44].

### 2D k spectrum evolution of the magnetic fields

As previously explained, the Weibel instability starts growing with many different wavelengths, and thus a $k$-resolved analysis is necessary for understanding the instability evolution. In Fig. 3 we show the 2D $k$-spectrum of the retrieved magnetic fields ($B_x$, Fig. 3a, and $B_y$, Fig. 3b). Here each $k$-spectrum was obtained by taking the 2D Fourier transform of the retrieved magnetic fields. These results highlight the evolution of the unstable modes and the transition to the dominating wavevector in the 2D $k$-space. For instance, we can see that both the $B_x$ and $B_y$ components start growing with a broad spectrum then the unstable region continuously shrinks in size and eventually narrow peaks appear which corresponds to quasi-single mode formation (marked by the white circles and arrows). For instance, in the last two frames (90 and 116.7 ps) of the $B_y$ spectra, a narrow peak with $\frac{k_x}{2\pi} \approx 0.007 \; \mu m^{-1}$ which corresponds to a structure with a wavelength of $\lambda_x \sim 145 \pm 20 \; \mu m$ and $\frac{k_y}{2\pi} \approx 0$ has formed.

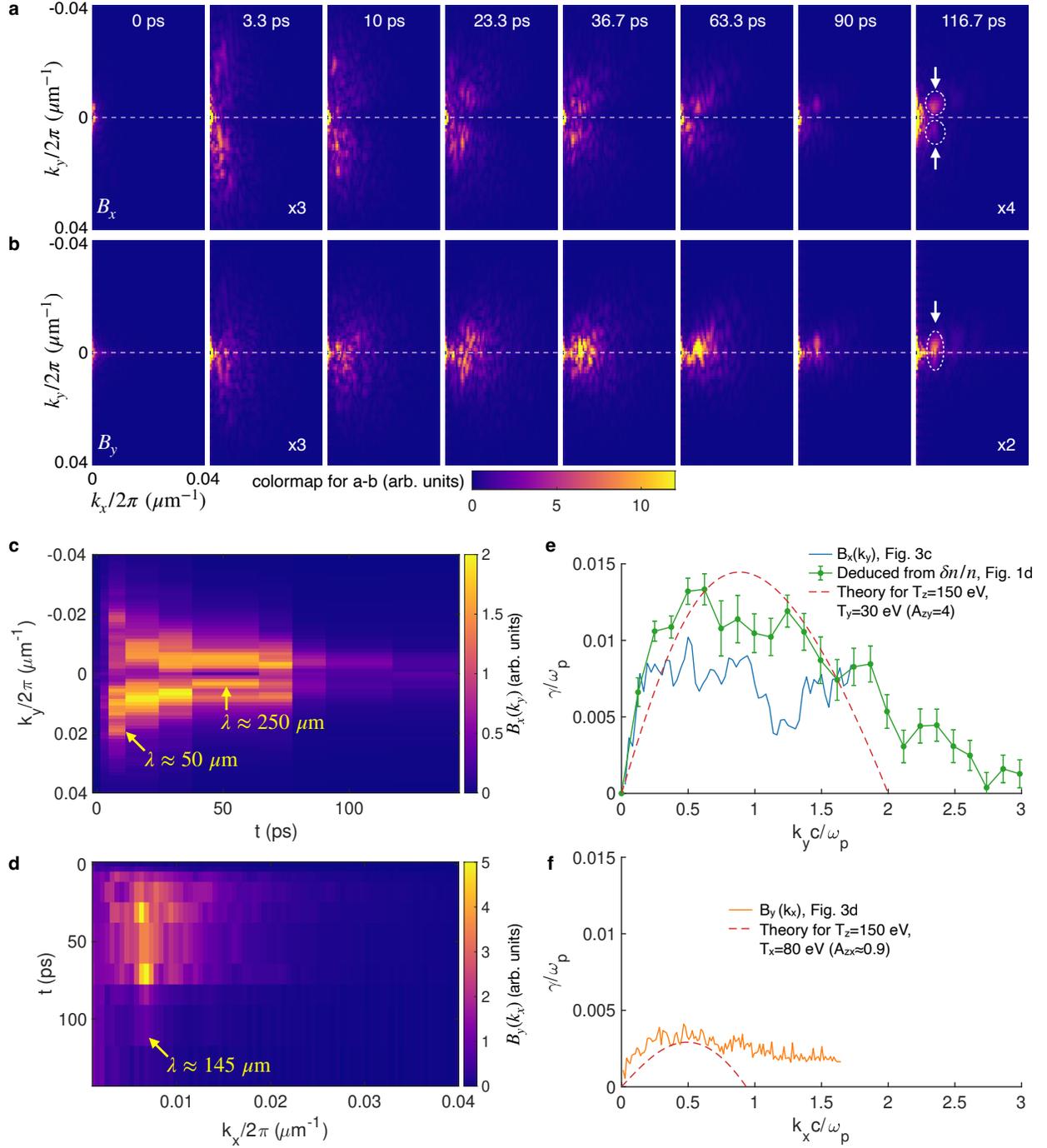

Fig. 3. *k*-spectrum evolution of the retrieved magnetic fields from the experimental data. (a)-(b) Evolution of the 2D *k*-spectrum for $B_x$ and $B_y$, respectively. The color scale for each frame is the same. The white circles and arrows mark the surviving quasi-single mode in the last frame. (c) Temporal evolution of the $k_y$ component of $B_x$ (averaged over $\frac{k_x}{2\pi} \in [0.002, 0.02]\ \mu m^{-1}$). The yellow arrow marks the dominant mode at different delays. (d) Evolution of the $k_x$ component of $B_y$ (averaged over $\frac{k_y}{2\pi} \in [-0.02, 0.02]\ \mu m^{-1}$). The yellow arrow marks the wavelength of the surviving quasi-single mode. (e) and (f), *k*-resolved growth rates and comparison with 1D kinetic theory. The blue curve in (e) shows the growth

rate deduced using the data in (d) and the green curve is deduced from the probe density modulation directly (the 3.3 ps frame in Fig. 1d). The relative peak amplitudes of the blue and the green curves obtained from the data are arbitrary (see supplemental). The orange curve in (f) is deduced using the data in (d). The dashed lines in (e) and (f) show the kinetic theory predictions.

The temporal evolution of the $B_x$ and $B_y$ spectra are plotted in Fig. 3c and d, respectively. The $B_x$ component begins with a spectral peak at $\lambda_y \approx 50 \ \mu m$ immediately following the onset of the instability (at 3.3 ps) that continuously shifts towards smaller $k_y$ or increasing wavelength. For instance, the wavelength of $B_x$ has increased to ~250 μm at ~50 ps and remains almost unchanged after that. This may be because the plasma is bounded in the transverse direction and therefore sets an upper limit for the wavelength. The $B_y$ field also starts with a broad spectrum following the onset of the instability, but its wavelength converges to ~145 μm at ~30 ps and remains almost unchanged for up to ~100 ps. We note that, however, in the experiment the wavelength of the $B_y$ field did continue increasing with time and reached ~300 μm at ~0.5 ns.

From the $k$- and time-resolved data shown in Fig. 3c and d we can deduce the $k$-resolved growth rates of the two magnetic field components. For instance, each row in Fig. 3c represents the temporal evolution of a specific $k_y$ component of the measured $B_x$ field, and therefore, the $k$-resolved growth rate is deduced by assuming an exponential growth and fitting the data (using the 0-3.3 ps data when the field grows most rapidly, see Supplemental Material). The result is shown by the blue curve in Fig. 3e. The growth rate peaks at $k_x \approx 0.5\omega_p c^{-1}$. Because the probe electron density modulation caused by the $B_x$ field appears as strips along the horizontal direction, which is equivalent to the time axis, it is also possible to estimate the growth rate using a single frame (e.g., the 3.3 ps frame in Fig. 1d) by tracking the increase of density modulation magnitude of each column from right to left which corresponds to an increasing delay (see Supplemental Material). The growth rate deduced from the density modulation is shown by the green curve in Fig. 3e. The two methods qualitatively agree with each other and the latter intraframe method gives a slightly larger growth rate due to higher temporal resolution. A similar analysis is applied to the $B_y$ field and the result is shown by the orange curve in Fig. 3f.

Using the tri-Maxwellian EVD obtained from PIC simulation, we can calculate the growth rate $\gamma(k)$ for the two magnetic field components that are perpendicular to the probe direction assuming no coupling between them in the linear phase (1D theory for each component, see Methods). The calculation results are shown by the red dashed lines in Figs. 3e and f. Reasonable agreement is seen for both field components, further supporting the argument that these fields are associated with the Weibel instability.

### Particle-in-cell simulation results

In this section, we present simulation results that qualitatively reproduce our experimental observations. The simulation was done using the Osiris 4.0 code[45] and it included both the ionization process and the subsequent multi-dimensional self-consistent evolution of the plasma. The details of the simulation can be found in the Methods section.

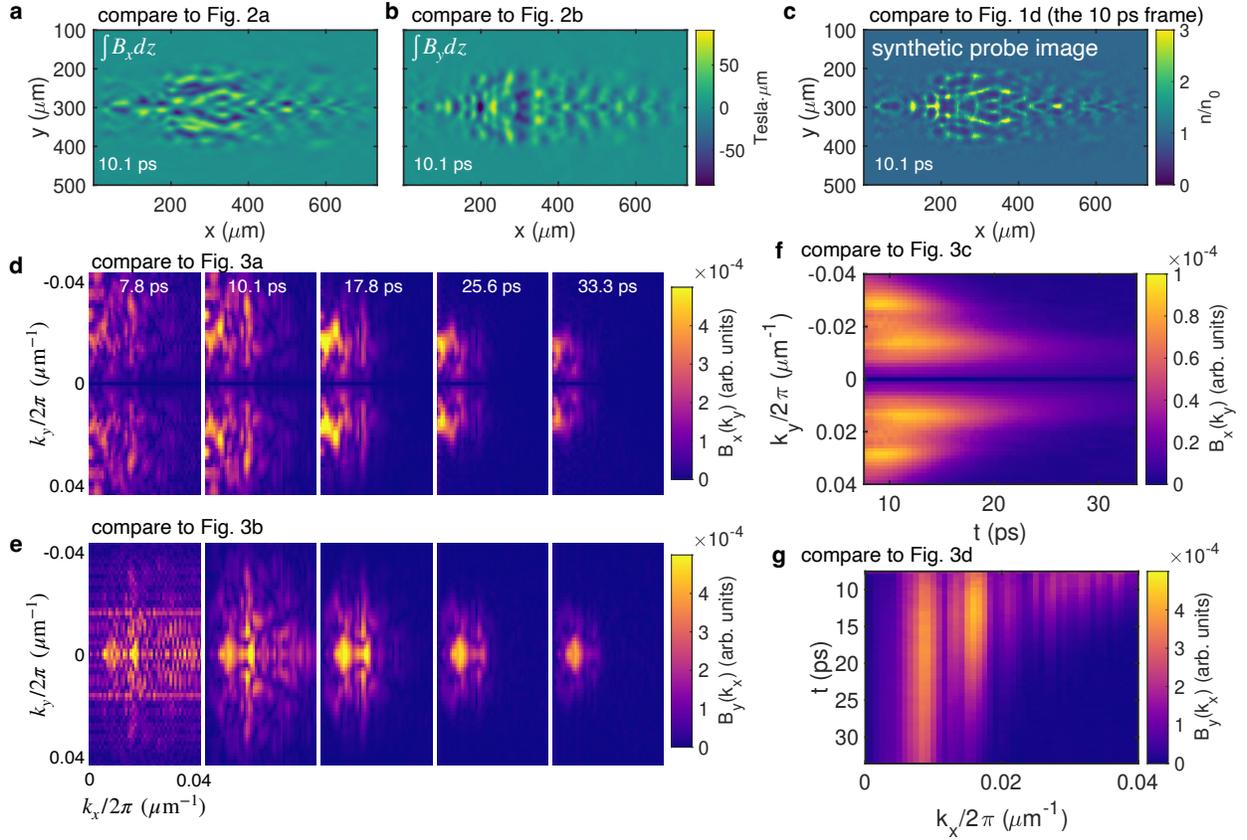

Fig. 4. Self-consistent 3D PIC simulation results. (a) and (b) show the path integrals of $B_x$ and $B_y$ obtained from the simulation. (c) is a synthetic probe image using the experimental parameters. (d) and (e) show the evolution of the 2D $k$-spectrum for path integrals of $B_x$ and $B_y$, respectively. (f) and (g) show the temporal evolution of the $k_y$ component of $B_x$ (averaged over $0.01 < k_x/2\pi(\mu m^{-1}) < 0.04$ ) and the $k_x$ component of $B_y$ (averaged over $-0.01 < k_y/2\pi(\mu m^{-1}) < 0.01$).

The simulation results are summarized in Fig. 4. The path integrals of the magnetic field components $B_x$ and $B_y$ are shown in Fig. 4a and b, respectively. Note that unlike in the experiment, here the integrations were taken by freezing the fields at an instantaneous time. Using these fields, a synthetic probe image was constructed and is shown in Fig. 4c. The image reproduces the observed qualitative features of the net structure. Several representative 2D $k$-spectra of the path-integrated magnetic fields ($\int B_x dz$ and $\int B_y dz$) are shown in Fig. 4d and e, respectively. As do the experimental plots shown in Figs. 3a and b, these plots also show the continuous narrowing of the initially broad 2D $k$ spectrum and the formation of a quasi-single mode at later times. The complete temporal evolution of the $k_y$ component of $B_x$ is shown in Fig. 4f and a similar plot for $B_y(k_x)$ is depicted in Fig. 4g. The key features observed in the experiment, including the transition to the dominant wavenumber in the 2D $k$-space and the narrowing of both the $k_y$- and $k_x$-spectra as a function of time, are also qualitatively reproduced. Although not shown, the RMS magnitude of the magnetic field reaches a peak of ~0.4 Tesla which is like that observed in the experiment and corresponds to about 0.9% of the thermal energy. Apart from these agreements, the instability seems to evolve about three times faster than that in the experiment. This may be due to artificial numerical collisions caused by only eight particles per cell to save simulation time.

***Discussion***

We have shown that the electron Weibel instability can self-generate quasistatic magnetic fields in plasmas with temperature anisotropy induced by optical-field ionization by picosecond intense $CO_2$ laser pulses. The time-resolved measurements indicate that the magnetic field magnitude and spectrum evolve on a time scale of tens of ps which implies that the electron velocity distributions also evolve toward a thermal plasma on a similar time scale. Although the Weibel instability is thought to be purely electromagnetic, in the nonlinear stage ion motion may lead to density fluctuations that need to be measured using optical Thomson scattering.

The actual EVDs of the plasma in this experiment are anisotropic in all three dimensions which give rise to complex Weibel magnetic field structures. We have probed the fields along one direction and hence in the $xy$ plane. To get a complete characterization of the 3D distribution of the magnetic field, electron probing at multiple angles is necessary. A practical way for doing this could be rotating the polarization direction of the $CO_2$ laser.

In summary, we have presented an unambiguous existence proof of the electron Weibel magnetic fields driven by a large effective temperature anisotropy as originally envisioned by Weibel for the first time. From these measurements, we have retrieved the 2D distribution of the magnetic fields and plasma current density, deduced the ultrafast dynamics of the fields and current density including the evolution of their 2D $k$-spectra and the $k$-resolved growth rate, which agree with kinetic theory predictions and PIC simulations. We observed the formation of a quasi-single mode of the magnetic field after the instability saturates which can last for up to half a nanosecond. Upon saturation, the current density modulation magnitude reaches ~5% and about 1% of the thermal energy of the plasma is converted to the magnetic field energy. These results represent a significant advance of the experimental understanding of Weibel instability. In our opinion, the OFI platform used here has a great potential for exploring ultrafast magnetic field dynamics in relativistic, anisotropic plasmas produced by ionization of high Z atoms in the laboratory relevant to astrophysical plasmas.

## Methods

### *Experimental setup*

The experiment was carried out at the Accelerator Test Facility (ATF) of Brookhaven National Laboratory. The high-power $CO_2$ laser system is capable to deliver up to 5 TW power within a pulse duration of about 2 ps (full width at half maxima, FWHM)[32]. For the data presented here, the laser energy was kept at a few hundred mJ such that the laser was able to fully ionize the hydrogen gas jet without driving large-amplitude self-modulated wakes. The laser was focused by an F/2 off-axis parabola with a 3-mm hole and the focal spot size was measured to be $w_0 \approx 45$ μm in vacuum. To enlarge the transverse extent of the plasma, the gas jet was shifted towards the upstream of the laser (closer to the off-axis parabola) by about 2 mm. At this location, the laser spot size was calculated to be $w \approx 140$ μm. The laser was linearly polarized in the horizontal plane with its electric field pointing along the electron probe direction. The plasma was produced by ionizing a supersonic hydrogen gas jet emanating from a converging-diverging nozzle with an opening diameter of 5 mm. The laser beam axis was placed ~2.5 mm above the nozzle exit. For the data presented here, the backing pressure of the nozzle was ~87 psig. The neutral density profile of the gas jet was measured offline using our recently developed ionization induced plasma grating method[46] (see Supplemental Material). The peak plasma density, assuming the gas jet is fully ionized by the $CO_2$ laser, is $(1.8 \pm 0.2) \times 10^{18}$ cm$^{-3}$. As they traverse through the plasma, the probe electrons receive transverse deflections by the magnetic fields. Then the probe propagates in the vacuum and the angular deflection translates into electron flux modulation. A set of permanent magnet quadrupoles (PMQs) like those used in Refs. [34,35] was used to relay and magnify the electron probe image to a scintillator screen (100-μm thick YAG: Ce crystal with both sides polished and one side coated with indium tin oxide) that was placed ~0.5 meters away from the plasma. The electron flux distributions were converted to optical images by the scintillator crystal and the latter was further magnified using an objective (Mituyoto 5x Plan APO) and recorded by a charge-coupled device (Basler acA1920-50gm with 5.86 μm pixel size) equipped with a camera lens with remote focus control. The probe electron beam was delivered by the ATF linac at 1.5 Hz but the data was taken at a lower laser repetition rate using a trigger signal to synchronize the gas jet, $CO_2$ laser, e$^-$ beam, and the camera. The energy of the probe beam was fixed at 50.5 MeV (0.2% energy spread) to match the designed optimal energy for the PMQs. The pulse duration of the probe beam was tuned to be ~1-2 ps by varying the slit size to select a portion of the longer pulse. Although the emittance of the electron beam was not measured during this run, it was estimated to be ~1.4 mm·mrad based on previous measurements[33]. The focus of the electron beam was shifted upstream such that the beam was able to cover a few mm field of view.

### *Electron imaging system*

The key part of the electron imaging system is the PMQs that can relay the electron images at the object plane to the image plane by a 2D one-to-one mapping. The set of PMQs includes four pieces (two identical pairs) of Halbach-type high-gradient quadrupoles arranged in the A-B-B-A order. Two pairs of the PMQs were mounted separately such that the separation between them can be adjusted to change the magnification and move the object plane location. The two assemblies were held by a third motorized translation stage which allows the PMQs to be moved away from the electron beam path. In the PMQ-in configuration, the electron probe formed images at $10 \pm 0.5$ mm (tunable from 0 to 25 mm) from the plasma, and then the image was relayed and magnified by the PMQs. In the PMQ-out configuration, the electron probe formed images directly on the scintillator screen. In both configurations, the overall magnification and resolving power of the imaging system were calibrated using transmission electron microscope (TEM) grids. The magnifications were ~7.8x and ~3.7x and the resolving powers were ~2.9 µm and ~10 µm for the two configurations.

### *Retrieval of magnetic fields*

To retrieve the magnetic fields, the first step was to solve for the deflection angles $\alpha_x$ and $\alpha_y$ of the probe electrons. Using the measured probe density profile $n(x, y)$, a synthetic background $n_0$ was estimated by smoothing the image using a code based on a penalized least squares method[47]. Then the relative density modulation was calculated as $n/n_0 - 1$. The algorithm in Ref. [39] was used to calculate the optimal transport of a uniform background to the relative density modulation. Before doing the calculation, the region of interest of the image was selected and down-sampled by a factor of 10 to reduce its size to $90 \times 150$ pixels to reduce the computational resource requirements to an affordable level. Then the deflection angles were used to calculate the transverse magnetic fields $B_x$ and $B_y$, with the assumption that the plasma has a slab geometry with a thickness of 300 µm. The current density $J_z$ was calculated using Ampere's law as explained in the main text.

### *Growth rates calculation*

The growth rates of the Weibel instability were calculated using 1D kinetic theory. The dispersion relation for the Weibel mode in an anisotropic plasma is[48]

$$1 - \frac{c^2 k^2}{\omega^2} + \frac{\omega_p^2}{\omega^2}[A + (A + 1)\xi Z(\xi)] = 0$$

where $\omega = i\gamma$ is the complex frequency, $\gamma$ is the growth rate, $k$ is the wavenumber of the magnetic field, $A = \frac{T_{\text{hot}}}{T_{\text{cold}}} - 1$ the temperature anisotropy, $\xi = \frac{\omega}{\sqrt{2}kv_{\text{hot}}}$, $v_{\text{hot}} = \sqrt{\frac{k_B T_{hot}}{m_e}}$ is the thermal velocity in the hot temperature direction, and $Z(\xi)$ is the plasma dispersion function. The theoretical growth rates in Fig. 3e and f were calculated using ($T_z = 150$ eV, $T_y = 30$ eV, $A_{zy} = \frac{T_z}{T_y} - 1 = 4$) and ($T_z = 150$ eV, $T_x = 80$ eV, $A_{zx} = \frac{T_z}{T_x} - 1 = 0.875$), respectively.

### *PIC simulation*

The 3D PIC simulation was performed using the Osiris 4.0 code[45]. The stationary simulation box has dimensions of $500(x) \times 400(y) \times 400(z) \, c/\omega_0$. Here $\omega_0$ is the frequency of the $CO_2$ laser ($\lambda = 9.2 \, \mu m$). A hydrogen (atomic) gas was initialized inside the box. In the laser propagation direction, the gas density profile contains two $10 \, c/\omega_0$ linear ramps on both ends and two $10 \, c/\omega_0$ gaps were set between the gas and box boundary. In the other two orthogonal directions, the gas density is uniform. The peak plasma density was set to $10^{18}$ cm$^{-3}$. A linearly polarized laser with the experimental parameters ($\lambda = 9.2 \, \mu m$, $\tau = 2$ ps, $w_0 = 45 \, \mu m$, $E = 115$ mJ) was launched from the left wall of the box. The vacuum focal plane of the laser was set to $x = 1520 \, c/\omega_0$, which means that the simulation box covers the region that is ~1.5 mm upstream of the focus to mimic the experimental condition. The ionization of hydrogen was calculated using the Ammosov-Delone-Krainov (ADK) model[49] and the subsequent evolution of the OFI plasma was self-consistently modeled. The 3D distribution of the magnetic fields and plasma currents were saved every ~0.5 ps for analysis.

## Acknowledgments

This work was supported by the U.S. Department of Energy Grant No. DE-SC0010064 and DE-SC0014043 as well as NSF Grant No. 1734315. The simulations were performed on the NERSC Cori cluster operated under Contract No. DE-AC02-5CH11231 at LBNL. The authors thank Dr. Mark Palmer, Prof. Vladimir Litvinenko and the ATF staff for their support and hospitality throughout this work. The authors also acknowledge Dr. Yu Fang for his help with measuring the PMQs.

## Author contributions

C.Z. and C.J. conceived the project. C.Z., Y.W., M.S., A.F., and K.M. designed and performed the experiments with the help from I.P., N.V.N., A.G., R.K., K.K., M.F., I.P. and M.P.. C.Z. and C.J. carried out the analysis and wrote the manuscript. All authors contributed extensively to the work presented in this paper.